\documentclass[12pt,reqno]{amsart}
\usepackage{cite}
\usepackage{amssymb,latexsym,amsmath,amsfonts}
\usepackage{latexsym}
\usepackage[mathscr]{eucal}
\usepackage[utf8]{inputenc}
\usepackage{palatino}
\usepackage{scalerel}
\usepackage{mathtools}
\numberwithin{equation}{section}

\newtheorem{thm}{Theorem}[section]
\newtheorem{defn}{Definition}[section]
\newtheorem{prop}{Proposition}[section]

\newtheorem{lemma}{Lemma}[section]

\begin{document}

\title[On LCP and checkable group codes]{On LCP and checkable group codes over finite non-commutative Frobenius rings}
\author[S. Bhowmick ]{Sanjit Bhowmick$^{1}$}
\address{$^{1}$ Sanjit Bhowmick,
                    Department of Mathematics,
                    National Institute of Technology  Durgapur,
                    West Bengal, India.}                                                  
                    \email{sanjitbhowmick392@gmail.com}
\author[J. de la Cruz]{Javier de la Cruz$^{2}$}
\address{$^{2}$ Javier de la Cruz,
	Departamento de Matem\'aticas, Universidad del Norte,   Barranquilla, Colombia}                                                  
\email{jdelacruz@uninorte.edu.co}
   \author[E. Martínez-Moro]{Edgar Martínez-Moro$^{3}$}
   \address{$^{3}$ Edgar Martínez-Moro,
 Institute of Mathematics, University of Valladolid, Castilla, Spain }
   \email{edgar.martinez@uva.es}   
   
   \author[Anuradha Sharma]{Anuradha Sharma$^{4}$}
   \address{$^{4}$ Anuradha Sharma,
   	Department of Mathematics, IIIT-Delhi, New Delhi- India
     }
   \email{anuradha@iiitd.ac.in}

%\thanks{This work is supported by a UGC-CSIR fellowship.}

\begin{abstract}
 We provide a simple  proof for a complementary pair of group codes over a finite non-commutative Frobenius ring of the fact that one of them is equivalent to the other one. We also explore this fact for checkeable codes over the same type of alphabet. 
\end{abstract}
\maketitle 
\section{Introduction}
A linear code $C$ of length $n$ over a finite ring $\mathcal R$ is just a submodule of $\mathcal R^n$. A pair of linear codes $(C,D)$ of the same length is called a \textit{linear complementary pair} (LCP) if they meet trivially, i.e. $C \cap D = \{ 0 \}$,  and they cover all the space $C + D = \mathcal R^n$. LCP of codes over finite fields % which are a class of unique characteristics, 
have been of interest and extensively explored because of their %complex 
algebraic structure and wide applications in cryptography. Carlet et al. \cite{Bringer2014} showed that LCP of codes can be used to improve the security of data processed by sensitive devices, particularly against so-called side-channel assaults and fault injection attacks. %Masking is the most generic and effective known protection against SCA: every sensitive data is bitwise added with a uniformly distributed random vector of the same length or multiple ones, referred to as a mask globally. It is possible to extract the sensitive data from the masked data if the sensitive data and the mask both belong to two supplemental subspaces, $C$ and $D$ of a larger vector space. it is proven that 
%and the level of resistance to both SCA and FIA is determined by the security parameter min$\{d(C),d(D^\perp)\}$, where $d(C)$ means minimum distance of $C$ and $D^\perp$ is the dual of $D$. 
In particular, the case one of the codes is the dual of the other one $C^\perp=D$, is \textit{called linear complementary dual} (LCD) code that was introduced by Massey \cite{Massey1992} and it has been extensively studied (see for example \cite{Bhowmick} and the references within). Recently, it has been shown that every linear code over a finite field of size greater than $3$ is equivalent to an LCD code \cite{CMTQP18}.

Group codes are an extension of the concept of cyclic codes and they can be seen as ideals in the group ring $\mathcal R G$, where $\mathcal R$ is a finite ring and $G$ a finite group. From the very begining there were stated conditions for linear and cyclic codes over a field to be LCD \cite{ Massey1992,Yang1994} and it was extended to group codes over finite fields in \cite{Cruz2017}. In \cite{Oz1},  the authors proved that for linear complementary pairs $(C,D)$, the codes $C$ and $D^\perp$ (the dual od $D$)  are equivalent if and are both cyclic or 2D-cyclic codes, and to $n$-cyclic in \cite{Oz2}, both papers worked their results under the assumption that the characteristic of the field does not divide the length, i.e. they where semisimple. This result was extended to any LCP of group codes over a finite field (despite its characteristic) in \cite{Bor2020}.  The result was further generalized for finite chain rings in \cite{Gu20} and recently to commutative Frobenius rings in \cite{LL}. In this paper, we will cover the final gap till non-commutative Frobenius rings that constitute the more general type of alphabet for coding theory \cite{Wood1999}. Despite we cover the same type of rings that \cite{LL} in the non-commutative case, our techniques differ from that ones and they mostly rely on the decomposition of modules and idempotents.

 Finally we will pay attention to chekeable codes in $\mathcal R G$. If we are dealing with $G$ an abelian code,  one important feature of semisimple codes over fields or rings is that all of them can be generated by a single codeword, i.e., they can be regarded as principal ideals.
 This property is not generally true in the modular case, and it partly explains
 the reason why those codes are more difficult to describe. However, that of
 all the ideals in  are principal is not equivalent to the semisimple condition, see for example \cite{Edgar} and the references therein for a polynomial description of Abelian codes over finite chain rings and  finite fields. In this paper we will deal with checkable codes in $\mathcal R G$, that is, ideals in $\mathcal R G$ whose anhihilator is a principal right ideal, see \cite{Bor2022} for a study of checkable group codes over finite fields.
 
 The outline of the paper will be as follows. Section~\ref{sec:pre} will show some preliminaries on non-commutative Frobenius rings and modules over them. For a given group ring $\mathcal R G$, where $G$ is a finite group and $\mathcal R$ a non-commutative Frobenius ring, Section~\ref{sec:idem}  will show its idempotents and the related decomposition of a $\mathcal R$ -module in $\mathcal{R}G$. Sections~\ref{sec:local}~\&~\ref{sec:gen} study when a group code is LCP in the case of the base ring being a  a non-commutative finite local ring or a non-commutative finite Frobenius ring respectively. Finally, in Section~\ref{sec:check}, we deal with checkable codes in $\mathcal R G$.

\section{Preliminaries}\label{sec:pre}
Throughout this paper, $\mathcal{R}$ will denote a finite Frobenius ring with identity $1_\mathcal{R}$. Furthermore,
unless mentioned otherwise, we assume our rings to be non-commutative.  In this section we will provide a brief overview to some results concerning modules over $\mathcal{R}$ (for a more detailed study see for example \cite{Dum04}) and some definitions and notations related to group codes.

A right $\mathcal{R}$-module $A$ is called free $\mathcal{R}$-module if $A$ is isomorphic to $\mathcal{R}^t$ as a $\mathcal{R}$-module, where $t\geq 1$.

\begin{defn}[Projective \& injective modules]$ $
	\begin{enumerate}
		\item A right $\mathcal{R}$-module $P$ is called projective if there exists another right $\mathcal{R}$-module $Q$ such that $P\oplus Q$ is free $\mathcal{R}$-module.
		\item A right $\mathcal{R}$-module $I$ is said to be injective if  for any monomorphism $g:A\rightarrow B$ of right $\mathcal{R}$-modules and any $\mathcal{R}$-homomorphism $h:A\rightarrow I$, there exists an $\mathcal{R}$-homomorphism $h':B\rightarrow I$ such that $h=h'\circ g$. 
	\end{enumerate}
\end{defn}
The following results can be also found in \cite{Dum04}.
\begin{thm}\label{th-0.0001} $ $ 
	\begin{enumerate}
		\item (Baer's criterion) A right $\mathcal{R}$-module $I$ is injective if and only if, for any right ideal $\mathcal I$ of $\mathcal{R}$, any $\mathcal{R}$-homomorphism $\gamma:\mathcal I\rightarrow I$  extends to $\mathcal{R}\rightarrow I$, that is, $\gamma=c\cdot$ is the multiplication by an element $c\in I$.
\item $P=\bigoplus\limits_{i=1}^{n} P_i$ is projective $\mathcal{R}$-module if and only if  $P_i$ is projective $\mathcal{R}$ module for each $i\in\{1,2,\dots,n\}$.
\item $I=\bigoplus\limits_{i=1}^{n} I_i$ is injective $\mathcal{R}$-module if and only if each $I_i$ is injective $\mathcal{R}$ module for each $i\in\{1,2,\dots,n\}$.
\end{enumerate}
\end{thm}
A finite ring $\mathcal{R}$ is said to be Frobenius if $\mathcal{R}$ forms a self-injective right $\mathcal{R}$-module (left $\mathcal{R}$-module). Alternatively, for a given finite ring $\mathcal{R}$ is said to be Frobenius if right $\mathcal{R}$-module $\dfrac{\mathcal{R}}{J(\mathcal{R})}$ is isomorphic to $\mathrm{Soc}(\mathcal{R})$ as a right $\mathcal{R}$-module, where $J(\mathcal{R})$ denotes the intersection of all maximal ideals in $\mathcal{R}$ and $\mathrm{Soc}(\mathcal{R})$ the socle of $\mathcal{R}$, which is the sum of all irreducible ideals in $\mathcal{R}$ (for more details see \cite{Lam19}). 

\begin{thm}\label{th-0.02} The following statements are equivalent:
\begin{itemize}
\item[(1)] $\mathcal{R}$ is Frobenius,
\item[(2)] every ideal in $\mathcal{R}$ is projective if and only if it is injective in $\mathcal{R}$.
\end{itemize}
\end{thm}

A nonempty subset $C\subseteq \mathcal{R}^n$ is called a left (right) linear code of length $n$ over $\mathcal{R}$ if it is an left (right) $\mathcal{R}$-submodule of $\mathcal{R}^n$. In this paper all codes are assumed to be linear. We will denote by  $G$   a finite group  and by $\mathcal{R}G$   the collection of formal sums
$\{\sum_{g\in G}a_g g : g \in G, a_g \in \mathcal{R} \},$
with the componentwise addition and the product given by $a \cdot b = ab =\sum_{g\in G}\left(\sum_{h\in H}a_hb_{h^{-1}g}h \right)g, $
for $a =\sum_{g\in G}a_g g$ and $b =\sum_{g\in G}b_g g$.
Under this operations,  $(\mathcal{R}G, +, \cdot)$ becomes an $\mathcal{R}$-module with the elements of $G$ as a basis and therefore we can think the ideals in $\mathcal{R}G$ as codes in $\mathcal{R}^{|G|}$. That is,  a  right (left) group code $C$ in $\mathcal{R}G$  is a right (left) ideal in $\mathcal{R}G$, which is indeed an $\mathcal R$-linear code if we see an element  $a =\sum_{g\in G}a_g g$ as $(a_g)_{g\in G}\in \mathcal R^{|G|}$ for a given ordering on the elements in $G$. Throughout this paper, by a group code $C$, we will  always mean a right ideal unless other thing is stated. % Note also that  $(\mathcal{R}G, +, \cdot)$  is a ring containing $\mathcal{R}$ in its center if $\mathcal{R}$ is commutative. 
%Therefore, an $\mathcal{R}$-algebra structure is endowed into $\mathcal{R}G$ via multiplication in $G$. 

For an element $a =\sum_{g\in G}a_g g\in\mathcal{R}G$, the adjoint of $a$ is defined as 
$\widehat{a} =\sum_{g\in G}a_g g^{-1}$, and it is called self-adjoint if $a = \widehat{a}$. We will denote by  $Z(\mathcal{R}G)$  the center of  $\mathcal{R}G$.

The group ring $\mathcal{R}G$ has a symmetric non-degenerate $G$-invariant bilinear form $\langle~,\rangle$ defined for the elemnets in $G$ as 
$$ \langle g,h\rangle = \left\{
\begin{array}{lll} 1 &  \mbox{if}~g=h,\\ 
0 &  \mbox{otherwise}\\
 \end{array} \right. ,$$ that can be straightfoward extended to  $\mathcal{R}G$.
 Note that the  $G$-invariant means $\langle ag,bg\rangle=\langle a,b\rangle$ for all $a,b$ in $\mathcal{R}G$ and $g$ in $G$. The above form relates to the Euclidean inner product via the isomorphism $\mathcal{R}G$ and $\mathcal{R}^{|G|}$. With respect to this form, we can define the dual code $C^\perp$ of a group code $C$ as usual  $C^\perp=\{ a\in \mathcal{R}G\mid \langle a,c\rangle= 0\hbox{ for all } c\in C \}$.
 
 As usual,  if $C_1$ and $C_2$ are two codes in $\mathcal{R}G$, then
  $(C_1+C_1)^\perp=C_1^\perp\cap C_2^\perp$ and 
   $C_1^\perp+C_2^\perp=(C_1\cap C_2)^\perp$. It is also well known that if
 	 $C$ be a group code in $\mathcal{R}G$, then 
 	$\mid C\mid\mid C^{\perp}\mid=\mid\mathcal{R}G\mid$
(it follows from J. Wood's result in \cite{Wood1999}, since $\mathcal{R}G$ is a finite Frobenius ring).

 Let $C$ and $D$ be two linear codes of length $n$ over $\mathcal{R}$, we say that the pair $(C,D)$ is a linear complementary pair (LCP) if $C\oplus D=\mathcal{R}^n$. Analogously,
  a pair of codes $(C,D)$ in $\mathcal{R}G$, is called an LCP group code  if $C\oplus D=\mathcal{R}G$. 

\section{Idempotents and the decomposition of $\mathcal{R}G$-modules}\label{sec:idem}

In this section we will review some results related to the relationship between the idempotents of the ring  $\mathcal{R}G$ and 
 the decomposition of right $\mathcal{R}G$-submodules. The results are in correspondence with those in \cite{Cruz2017}  where the finite field case was studied.
 
  Let $M$ be a right $\mathcal{R}G$-submodule in $\mathcal{R}G$, then the dual module $M$ is denoted by $M^\star$ and defined by $M^\star =\{f~|~f:M\rightarrow \mathcal{R}\hbox{ is ~an ~}\mathcal{R} \hbox{~module~homomorphism}\}.$
    $M^\star$ is an $\mathcal{R}G$-right submodule under the action
 $$(fg)(m)=f(mg^{-1})~{ m\in M,~f\in M^{\star},~g\in G},$$
 and clearly, $|M|=|M^\star|.$
 \begin{lemma}
 If $e$ is an idempotent element in $\mathcal{R}G$, then $\widehat{e}\mathcal{R}G\cong e\mathcal{R}G^\star$.
 \end{lemma}
 \begin{proof} Consider the mapping $\varphi:\widehat{e}\mathcal{R}G\rightarrow eRG^\star$ by $\widehat{e}r\mapsto \varphi_{\widehat{e}r}$, with $$\varphi_{\widehat{e}r}(es)=\langle es, \widehat{e}s\rangle, ~where~r,s\in \mathcal{R}G.$$ It is clear that $\varphi$ is well define and that $\varphi$ is group homomorphism under addition, as $\varphi(\widehat{e}r+\widehat{e}s)=\varphi(\widehat{e}r)+\varphi(\widehat{e}s)$, for all $r,s\in \mathcal{R}G$. Next, for any $\theta\in\mathcal{R}G$ and $\widehat{e}r$ in $\mathcal{R}G$, we have to prove that $\varphi(\widehat{e}r\theta)=\varphi(\widehat{e}r)\theta$. Since 
 $$\varphi_{\widehat{e}r\theta}(es)=\langle es, \widehat{e}r\theta\rangle=\langle es\widehat{\theta},\widehat{e}r\rangle=\varphi_{\widehat{e}r}(es\widehat{\theta})=(\varphi_{\widehat{e}r}\theta)(es).$$ we have that  $\varphi(\widehat{e}r\theta)=\varphi(\widehat{e}r)\theta$, where $\theta\in \mathcal{R}G$. Thus, it follows that $\varphi$ is a right $\mathcal{R}G$-module homomorphism. Note that the kernel of $\varphi$ is given by
 \begin{eqnarray*}
ker(\varphi)&=&\{\widehat{e}r~|~\varphi(\widehat{e}r)=0\}, \\
            &=&\{\widehat{e}r~|~\varphi_{\widehat{e}r}(es)=0,~for~all~s\in\mathcal{R}G\}, \\
            &=&\{\widehat{e}r~|~\langle es, \widehat{e}r\rangle=0,~for~all~s\in\mathcal{R}G\} = \{0\}.
\end{eqnarray*} 
Therefore, $\varphi$ is a monomorphism. Since $|\hat{e}\mathcal{R}G|=|e\mathcal{R}G|=|eRG^\star|$, it follows that $\varphi$ is onto. Thus, $\varphi$ is a left $\mathcal{R}G$-module isomorphism.
 \end{proof}

\begin{thm}\label{thm-0.1}
Let $\mathcal{R}G$ be a finite group ring with $1_{\mathcal{R}G}$. Then there is a one-one correspondence between the following statements:
\begin{itemize}
\item[(1)] There is a decomposition $\mathcal{R}G=I_1\oplus\cdots\oplus I_s$ as a direct sum of right $\mathcal{R}G$-submodules,
\item[(2)] there exist orthogonal idempotents $e_1$, $\dots$, $e_s$ such that $e_1+\cdots+e_s=1_{\mathcal{R}G}$. 
\end{itemize}
\end{thm} 
\begin{proof}
Let us assume that $\mathcal{R}G=I_1\oplus I_2\oplus\cdots\oplus I_s$. Since $1_{\mathcal{R}G}$ in $\mathcal{R}G$, we have that $1_{\mathcal{R}G}=\sum_{i=1}^sx_i$, where $x_i\in I_i$ for all $1\leq i\leq s$. Thus

 \begin{eqnarray*}
         x_i&=&1\cdot x_i ={\sum_{j=1}^s}x_jx_i \\
   x_i-x^2_i&=& \sum_{j(\neq i)=1}^sx_jx_i\in I_i\cap (I_1+\dots+ I_{i-1}+I_{i+1}+\dots +I_{s}), 
\end{eqnarray*}
and  $x_i-x_i^2=0$ for all $1\leq i\leq s$ and $x_ix_j=0$ for all $i\neq j$ as required.

Conversely, given a set of orthogonal idempotents $\{e_i\}$, let us define $I_i=e_i\mathcal{R}G$ which is a right submodule of $\mathcal{R}G$. Since $\sum\limits_{i=1}^se_i=1_{\mathcal{R}G}$, this implies $\mathcal{R}G=I_1+I_2+\cdots+I_s$. Therefore, $\mathcal{R}G=\bigoplus\limits_{i=1}^se_i\mathcal{R}G$, as $e_i$ is an orthogonal idempotent element in $\mathcal{R}G$.
\end{proof}
\begin{thm}\label{thm-0.2}
Let $\mathcal{R}G$ be a group ring with identity $1_{\mathcal{R}G}$ and $e$ an idempotent in $\mathcal{R}G$. Then $e\mathcal{R}G$ is indecomposable if and only if $e$ is primitive.
\end{thm}
\begin{proof}
Let us assume $e\mathcal{R}G$ is indecomposable, therefore , let $e=e_1+e_2$, where the $e_i$ are orthogonal idempotent elements. First, let $x\in e_1\mathcal{R}G\cap e_2\mathcal{R}G$, this implies $$x=e_1^2r_1=e_1e_2r_2=0,~ \mbox{for ~some}~r_1,r_2\in\mathcal{R}G.$$ It follows that $e_1\mathcal{R}G\cap\mathcal{R}G=\{0\}$, clearly $(e_1+e_2)\mathcal{R}G\subset e_1\mathcal{R}G\oplus e_2\mathcal{R}G$.

Let $e_1s_1+e_2s_2\in e_1\mathcal{R}G\oplus e_2\mathcal{R}G$ for some $s_1, s_2\in\mathcal{R}G$.
Now, $$(e_1+e_2)(e_1s_1+e_2s_2)\in(e_1+e_2)\mathcal{R}G.$$
Hence, $e\mathcal{R}G=(e_1+e_2)\mathcal{R}G=e_1\mathcal{R}G\oplus e_2\mathcal{R}G$.

Conversely, if $e\mathcal{R}G=I+J$, express $e=x+y$, where $x\in I$ and $y\in J$.

Since $I\subseteq e\mathcal{R}G$, then $x=er_1$, for some $r_1\in \mathcal{R}G$. Therefore, $$ex=e^2r_1=er_1=x.$$
It follows that $x=ex=x^2+yx$, i.e., $x-x^2=yx\in I\cap J$, that means $x=x^2$ and $yx=0$. Similarly, $y=y^2$ and $xy=0$,which is a contradiction. 
\end{proof}
\begin{thm}\label{thm-0.3}
If $\mathcal{R}G$ is a finite module with identity $1_{\mathcal{R}G}$, then $\mathcal{R}G$ can be written as direct sum of indecomposable submodules.
\end{thm}
\begin{proof}
%If possible, let $\mathcal{R}G$ can not be written as direct sum of indecomposable, i.e., $\mathcal{R}G=M_1\oplus A$, where $A$ is not a finite direct sum of indecomposable. Again $A=M_2\oplus A_1$, where $A_2$ is not a finite direct sum of indecomposable. Continuing this process, we get after $i$-th processes $$\mathcal{R}G=M_1\oplus M_2\oplus\cdots\oplus M_i\oplus A_{i-1},$$ where $A_{i-1}$ is not a finite direct sum of indecomposable.
%The induction proceeds to give us an ascending chain
%$$0<M_1<M_1\oplus M_2<\cdots<M_1\oplus\cdots\oplus M_s<\cdots$$
%of submodules.
It follows directly from the contradiction between an ascending chain of decompositions and the fact that
 $\mathcal{R}G$ is finite.
\end{proof}

\section{LCP group codes over finite non-commutative local rings}\label{sec:local}
In this section, we will  assume that $\mathcal{R}$ is a finite non-commutative local ring, that is, whenever $x, y\in \mathcal{R}$ are non-unit, so is $x+y$. Let  $\mathfrak{m}=J(\mathcal R)$ the Jacobson radical  of $\mathcal{R}$, it is well known that it is the set of non-units (since in an artinian local ring, every element is either unit or nilpotent). By Wedderburn's little theorem   every finite domain is a field, thus we will denote by  $\mathbb{F}_q=\mathcal{R}/\mathfrak{m}$   its  residue field. There is a natural homomorphism from $\mathcal{R}$ onto $\mathbb{F}_q$, i.e.,
$\pi:\mathcal{R}\rightarrow \mathcal{R}/\mathfrak{m}=\mathbb{F}_q,~r\mapsto\pi(r)=r+\mathfrak m ,~\mbox{for~any}~r~\mbox{in}~\mathcal{R}.$

This natural homomorphism from $\mathcal{R}$ onto $\mathbb{F}_q$ can be extended naturally to a homomorphism from $\mathcal{R}^n$ onto $\mathbb{F}^n_q$. It is clear that this map is a surjective ring homomorphism. Note that $\pi$ maps a linear codes over $\mathcal{R}$ to a linear code over $\mathbb{F}_q$.

%https://mathstrek.blog/2015/01/15/local-rings/

\begin{thm}\label{Thm-1.1}
If $C$ and $D$ are two right ideal (or right $\mathcal{R}G$-submodule) in $\mathcal{R}G$. Then the followings are equivalent
\begin{itemize}
\item[(1)] the pair $(C,D)$ is LCP in $\mathcal{R}G$;
\item[(2)] $C=e\mathcal{R}G$ and $D=(1-e)\mathcal{R}G$, where, $e=e^2\in \mathcal{R}G$.
\end{itemize}
\end{thm}
\begin{proof}
First, suppose that $(1)$ holds. Let $C+ D=\mathcal{R}G$ with $C\cap D=\{0\}$ and write $1=c+d$ with $c\in C$ and $d\in D$. It follows that 
          $c=c^2-dc=c^2-cd.$ Hence $c-c^2=cd=dc\in C\cap D$, thus $c=c^2$ and $cd=dc=0$. Furthermore, if we set $e=c$, then $e\mathcal{R}G\subset C$, $(1-e)\mathcal{R}G\subset D$ and, since  $C\oplus D=\mathcal{R}G=e\mathcal{R}G\oplus(1-e)\mathcal{R}G$, we have that $C=e\mathcal{R}G$ and $D=(1-e)\mathcal{R}G$ with $e=e^2$.
          
  Conversely, if we  suppose that $(2)$ holds, then $C=e\mathcal{R}G$ and $D=(1-e)\mathcal{R}G$ with $e=e^2\in \mathcal{R}G$. Since $e$ is an idempotent, we have $\mathcal{R}G=e\mathcal{R}G\oplus(1-e)\mathcal{R}G$. Therefore, we get the desired result.
\end{proof}
\begin{lemma}[See \cite{mil01}]\label{l-1.01}
If $I$ is an two-sided ideal in the ring $\mathcal{R}$ and $G$ is a finite group, then $\dfrac{\mathcal{R}G}{IG}\cong\dfrac{\mathcal{R}}{I}G$.
\end{lemma}
\begin{lemma}\label{l-1.02}
If $e$ is an idempotent element in $\mathbb{F}_qG$, then there exist an element $e'$ in $\mathcal{R}G$ such that $\pi(e')=e$.
\end{lemma}
\begin{proof}
Since $\mathfrak{m}=J(\mathcal R)$ is a two-sided ideal in $\mathcal R$, then by Lemma \ref{l-1.01}, we get $$\dfrac{\mathcal{R}G}{\mathfrak{m} G}\cong\dfrac{\mathcal{R}}{\mathfrak{m}}G\cong\mathbb{F}_qG$$ and $\dfrac{\mathcal{R}G}{\mathfrak{m}^i G}\cong\dfrac{\mathcal{R}}{\mathfrak{m}^i}G$, where $1\leq i \leq f$, and $\mathfrak{m}^f=0$ but $\mathfrak{m}^{f-1}\neq 0$ (note that in an artinian ring, the Jacoson radical is nilpotent). 

We will prove the result by induction. Let us define idempotent $e_i\in\dfrac{\mathcal{R}G}{\mathfrak{m}^iG}.$ as follows
            \begin{enumerate}
\item If $i=1$,  we set $e_1=e$, then this lemma holds.

\item If $i>1$, let $e_{i-1}$ is an idempotent element in $\dfrac{\mathcal{R}G}{\mathfrak{m}^{i-1}G}$. Let us consider $h$ be an element in $\dfrac{\mathcal{R}G}{\mathfrak{m}^iG}$ with its image $e_{i-1}$ in $\dfrac{\mathcal{R}G}{\mathfrak{m}^{i-1}G}$. 
\end{enumerate}
Then it is easy to verify that $h-h^2$ is in $\dfrac{\mathfrak{m}^{i-1}G}{\mathfrak{m}^iG}$. Hence, $(h-h^2)^2=0$.

Now, we set $e_i=(3h^2-2h^3)$.
Thus 
 $e_i^2-e_i = (3h^2-2h^3)(3h^2-2h^3-1) =(2h-3)(1+2h)(h-h^2)^2 =0.
$ and 
hhis lemma holds for all $i$. 
Since $\mathfrak{m}^f=0$, so $e'=e_f$.
\end{proof}
 \begin{thm}\label{Thm-1.1a}
 If $C$ and $D$ are two group codes in $\mathcal{R}G$ and $\pi(C)=e\mathbb{F}_qG$, $D=(1-e)\mathbb{F}_qG$ with $e=e^2$, then there exist an element $e'$ such that $C=e'\mathcal{R}G$, $D=(1-e')\mathcal{R}G$ with ${e'}^2=e'$ and
 $\pi(e')=e$.
 \end{thm}
 \begin{proof}
 Since $e$ is an idempotent element in $\mathbb{F}_qG$ with $\pi(C)=e\mathbb{F}_qG$, $\pi(D)=(1-e)\mathbb{F}_qG$, the pair $(\pi(C),\pi(D))$ is an LCP in $\mathbb{F}_qG$. Now from Lemma \ref{l-1.02}, we there exists $e'$ such that $\pi(e')=e$. In particular, take $C=e'\mathcal{R}G$ and $D=(1-e')\mathcal{R}G$.
 \end{proof}
 \begin{thm}
If $C$ and $D$ are two group codes over $\mathcal{R}G$, then the pair $(C,D)$ is a LCP in $\mathcal{R}G$ if and only if $(\pi(C),\pi(D))$ is a LCP in $\mathbb{F}_qG$.
\end{thm}
\begin{proof}
Let us assume that the pair $(C,D)$ is LCP in $\mathcal{R}G$, by the Theorem \ref{Thm-1.1}, there exist a idempotent $e$ in $\mathcal{R}G$ such that $C=e\mathcal{R}G,~~D=(1-e)\mathcal{R}G.$
Since $\pi$ is a surjective homomorphism from $\mathcal{R}G$ to $\mathbb{F}_qG$, then we obtain that 
  $$\pi(C)=\pi(e)\mathbb{F}_qG,~~\pi(D)=(1-\pi(e))\mathbb{F}_qG.$$
  Whence, the pair $(\pi(C),\pi(D))$ is a LCP in $\mathbb{F}_qG$.
  
Conversely, let us suppose that $(\pi(C),\pi(D))$ is a LCP in $\mathbb{F}_qG$. Then there exists an idempotent $e$ in $\mathbb{F}_qG$ such that $\pi(C)=e\mathbb{F}_qG$ and $\pi(D)=(1-e)\mathbb{F}_qG$. Then by Theorem \ref{Thm-1.1a}, there is $e'$ in $\mathcal{R}G$ such that $C=e'\mathcal{R}G$ and $D=(1-e')\mathcal{R}G$ with $e'^2=e'$. Hence the pair $(C,D)$ is a LCP in $\mathcal{R}G$.
 \end{proof}
 \begin{lemma}\label{lab-1}
If $C=e\mathcal{R}G$ with $e=e^2$, then the dual of $C$ is given by $C^{\perp}=(1-\widehat{e})\mathcal{R}G$. 
\end{lemma}
\begin{proof}
It is easily seen that $\widehat{e}^2=\widehat{e}$ and also $\langle ab,c\rangle=\langle b, \widehat{e}c\rangle$, for all $a,b,c$ in $\mathcal{R}G$. For any $x$ in $(1-\widehat{e})\mathcal{R}G$, then $x=(1-\widehat{e})f$, for some $f$ in $\mathcal{R}G$. Therefore, 
$\langle ea, x\rangle=\langle ea, (1-\widehat{e})f\rangle=\langle a, \widehat{e}(1-\widehat{e})f\rangle=0,$ for all $ea$ in $C$. Hence $(1-\widehat{e})\mathcal{R}G\subseteq C^{\perp}$. Taking into account that $C=e\mathcal{R}G$ and $e\mathcal{R}G\cong\widehat{e}\mathcal{R}G$    we have 
$$
 \mid C^{\perp}\mid =\dfrac{\mid \mathcal{R} G\mid}{\mid  C\mid} =\dfrac{\mid e\mathcal{R}G\mid\mid (1-e)\mathcal{R}G\mid}{\mid C\mid}= \mid (1-e)\mathcal{R}G\mid=\mid(1-\widehat{e})\mathcal{R}G\mid 
$$
Thus, we obtain the desired  result.
\end{proof} 
\begin{thm}
If $e\in Z(\mathcal{R}G)$ with $e=e^2$, then $C=e\mathcal{R}G$ and $D^\perp=\widehat{e}\mathcal{R}G$ are permutation equivalent. 
\end{thm}
\begin{proof}
Note that  $(C,D)$ is an LCP group code in $\mathcal{R}G$ with
$C=e\mathcal{R}G~\mbox{and}~D=(1-e)\mathcal{R}G,$ as $e$ is idempotent. By Lemma \ref{lab-1}  we know  that $D^{\perp}=\widehat{e}\mathcal{R}G.$ Thus for the group code $C$, $\widehat{C}$ is given by
\begin{align*}
 \widehat{C}& =\{\widehat{a}\in\mathcal{R}G~:~\forall~a\in C=e\mathcal{R}G\} =\{\widehat{e\theta}\in\mathcal{R}G~:~\forall~e\theta\in C=e\mathcal{R}G\}\\
 &= \{\widehat{\theta}\widehat{e}\in\mathcal{R}G~:~\forall~e\theta\in C=e\mathcal{R}G\} \stackrel{(*)}{=}\{\theta\widehat{e}\in\mathcal{R}G~:~\forall~\theta\in \mathcal{R}G\}\\
 &\stackrel{(**)}{=}\{\widehat{e}\theta\in\mathcal{R}G~:~\forall~\theta\in \mathcal{R}G\} =D^\perp,
\end{align*}
where $(*)$ holds because $RG\cong \widehat{RG}$ and $(**)$ holds because $e\in Z(\mathcal{R}G)$.
This completes the proof.
\end{proof}

\section{LCP group codes over finite non-commutative Frobenius rings}\label{sec:gen}
In this section, we will denote $\mathcal{R}$ as a finite non-commutative ring with identity $1_\mathcal{R}$. % We first recall some preliminaries which will have an essential role in this section.

%For a finite non-commutative ring $\mathcal{R}$ and any finite group $G$, we have $$\mathcal{R}G=\bigoplus\limits_{i=1}^{s}e_i\mathcal{R}G,$$ where $\sum\limits_{i=1}^{s}e_i=1$, $e_i^2=e_i$, for all $i=1,\dots, s$ and $e_ie_j=0$ for all $i\neq j$. 
%We start with the main result of this section.
\begin{thm}\label{th-5.1}
Let $C$ and $D$ be two group codes in $\mathcal{R}G$. Then the pair $(C,D)$ forms an LCP of group codes in $\mathcal{R}G$ if and only if $C=\oplus_{i=1}^ue_i\mathcal{R}G$ and $D=\oplus_{i=u+1}^{u+v}e_i\mathcal{R}G$ with $\sum_{i=1}^{u+v}e_i=1$, $e^2_i=e_i$ for all $1\leq i\leq u+v$ and $e_ie_j=0$ for all $i\neq j$.
\end{thm}
\begin{proof}
Let us assume that the pair $(C, D)$ is an LCP group code in $\mathcal{R}G$. Hence $C\oplus D=\mathcal{R}G$, so $C$ and $D$ both are projective. By Theorem \ref{thm-0.3}, $C$ can be written as the direct sum of indecomposable ideals, i.e.,
 $C=\bigoplus\limits_{i=1}^uI_i$, where $I_i=a_i\mathcal{R}G$ for all $1\leq i\leq u$.
 Since $C$ is projective and $\mathcal{R}G$ is Frobenius, then by Theorem \ref{th-0.02} we obtain that $I_i=a_i\mathcal{R}G$ is injective. Now, consider
 $
 \gamma :a_i\mathcal{R}G\rightarrow I_i$ given by $\gamma(a_ir)=a_ir$. By Baer's criterion in Theorem~\ref{th-0.0001} we have that $\gamma=m\cdot$, where $m\in I_i$, that is $m=a_ib$ with $b\in \mathcal{R}G$, and henceforth $a_i=\gamma(a_i)=ma_i=a_iba_i$, and $a_i\mathcal{R}G=e_i\mathcal{R}G$ where $e_i^2=e_i=a_ib$.
 Hence $a_i$ is idempotent element. Therefore, it follows from the Theorem \ref{thm-0.2} that  $a_i$ is primitive. Whence, $C=\bigoplus\limits_{i=1}^ua_i\mathcal{R}G$, where $a_i$ is primitive idempotent for all $1\leq i\leq u$.
 
 Similarly, $D=\bigoplus\limits_{j=1}^vb_j\mathcal{R}G$, where $b_i$ is primitive idempotent for all $1\leq i\leq u$. Since $C\oplus D=\mathcal{R}G$, then
 $$\left( \bigoplus\limits_{i=1}^ua_i\mathcal{R}G\right)\bigoplus\limits \left( \bigoplus\limits_{j=1}^vb_j\mathcal{R}G \right) =\mathcal{R}G.$$
 Then by Theorem \ref{thm-0.1}, we get set of idempotents $e_1,\dots,e_u,e_{u+1},\dots,e_{u+v}$ such that $C=\bigoplus\limits_{i=1}^ue_i\mathcal{R}G$ and $D=\bigoplus\limits_{i=u+1}^{u+v}e_i\mathcal{R}G$ with $\sum\limits_{i=1}^{u+v}e_i=1$ with $e^2_i=e_i$ for all $1\leq i\leq u+v$ and $e_ie_j=0$ for all $i\neq j$.
 
 Conversely, let $C=\bigoplus\limits_{i=1}^ue_i\mathcal{R}G$ and $D=\bigoplus\limits_{i=u+1}^{u+v}e_i\mathcal{R}G$ with $\sum\limits_{i=1}^{u+v}e_i=1$ with $e^2_i=e_i$ for all $1\leq i\leq u+v$ and $e_ie_j=0$ for all $i\neq j$. Therefore, $C+ D=\mathcal{R}G$, as $\sum\limits_{i=1}^{u+v}e_i=1$ and $C\cap D=\mathcal{R}G$ as $e_ie_j=0$ and $e^2_i=e_i$. Whence, the pair $(C,D)$ forms an LCP group code in $\mathcal{R}G$.
\end{proof}
\begin{thm}
Let $C$ and $D$ be two group codes in $\mathcal{R}G$. Then the pair $(C,D)$ forms an LCP group code in $\mathcal{R}G$ if and only if $C=e\mathcal{R}G$ and $D=(1-e)\mathcal{R}G$ with $e^2=e\in \mathcal{R}G$.
\end{thm}
\begin{proof}
In Theorem \ref{th-5.1}, we set $e=\sum_{i=1}^ue_i$. It is easily verified that $e^2=e$ and we get our required result.
\end{proof}
\begin{thm}
If $C=\bigoplus\limits_{i=1}^ue_i\mathcal{R}G$ with $\sum\limits_{i=1}^{u+v}e_i=1$, $e^2_i=e_i$ for all $1\leq i\leq u+v$ and $e_ie_j=0$ for all $i\neq j$, then the dual of $C$, i.e., $C^{\perp}=\bigoplus\limits_{i=u+1}^v\widehat{e_i}\mathcal{R}G$.
\end{thm}
\begin{proof} 
The dual of $C$, i.e.,
\begin{align*}
 C^\perp& =\left( \bigoplus\limits_{i=1}^ue_i\mathcal{R}G\right)^\perp  =\bigcap\limits_{i=1}^u(e_i\mathcal{R}G)^\perp \stackrel{(*)}{=} \bigcap\limits_{i=1}^u(1-\widehat{e_i})\mathcal{R}G\\
 &\stackrel{(**)}{=}(1-\sum\limits_{i=1}^u\widehat{e_i})\mathcal{R}G \stackrel{(***)}{=} \bigoplus\limits_{i=u+1}^v\widehat{e_i}\mathcal{R}G
 \end{align*}
where $(*)$ follows from {Lemma}~\ref{lab-1}, $(**)$ holds because  $e_ie_j=0$ and  $(***)$ because $\sum\limits_{i=1}^{u+v}e_i=1$.
 Thus, we get $C^{\perp}=\bigoplus\limits_{i=u+1}^v\widehat{e_i}\mathcal{R}G$.
 \end{proof}
\begin{thm}
If $e_i\in Z(\mathcal{R}G)$ with $\sum\limits_{i=1}^{u+v}e_i=1$, $e^2_i=e_i$ for all $1\leq i\leq u+v$ and $e_ie_j=0$ for all $i\neq j$, then $C=\bigoplus\limits_{i=1}^ue_i\mathcal{R}G$ and $D^\perp=\bigoplus\limits_{i=1}^u\widehat{e_i}\mathcal{R}G$ are permutation equivalent.
\end{thm}
\begin{proof} 
For given group code $C$,   $\widehat{C}$ is defined as follows (note that  $\widehat{}$ {~is ~homomorphism~under~addition} and an antisomorphism and that $e_i\in Z(\mathcal{R}G)$)
\begin{align*}
 \widehat{C} &=\left\{\widehat{c}~:~ c\in C=\bigoplus\limits_{i=1}^ue_i\mathcal{R}G \right\}  =\left\{\widehat{\sum\limits_{i=1}^ue_if_i}~:~ \sum\limits_{i=1}^ue_if_i\in\mathcal{R}G\right\} \\ &= \left\{\sum\limits_{i=1}^u\widehat{e_if_i}~:~f_i\in\mathcal{R}G\right\}  =\left\{\sum\limits_{i=1}^u\widehat{f_i}\widehat{e_i}~:~f_i\in\mathcal{R}G\right\}  =\left\{\sum\limits_{i=1}^u\widehat{e_i}\widehat{f_i}~:~f_i\in\mathcal{R}G\right\}\\ &=D^\perp,\end{align*}
 which completes the proof.
 \end{proof}
\section{Checkable codes}\label{sec:check}
In this section, we will characterize checkable codes in $\mathcal{R}G$ for a finite group $G$ over a finite Frobenius ring $\mathcal{R}.$  
Let $I$ be a right ideal in $\mathcal{R}G,$ the left annihilator of $I$ is defined by $\mathtt{Ann}_{l}(I)=\{r\in\mathcal{R}G~|~r\cdot a=0~\forall~a\in I\}.$ Let $J$ be a left ideal in $\mathcal{R}G,$ the right annihilator of $J$ is defined by $\mathtt{Ann}_{r}(J)=\{s\in\mathcal{R}G~|~a\cdot s=0~\forall~a\in J\}.$ 
The following result follows from \cite{Lam19,Wood1999}.

\begin{prop}\label{t1} For a group ring $\mathcal{R}G$ over Frobenius ring $\mathcal{R},$ then
\begin{enumerate} 
\item[(a)] $\mathtt{Ann}_{r}(a)=\mathtt{Ann}_{r}(\mathcal{R}Ga)$ for any $a$ in $\mathcal{R}G.$
\item[(b)] $\mathtt{Ann}_{l}(a)=\mathtt{Ann}_{l}(a\mathcal{R}G)$ for any $a$ in $\mathcal{R}G.$
\item[(c)] $I=\mathtt{Ann}_{l}(\mathtt{Ann}_{r}(I))$ for any left ideal $I$ in $\mathcal{R}G.$
\item[(d)] $J=\mathtt{Ann}_{r}(\mathtt{Ann}_{l}(J))$ for any right ideal $J$ in $\mathcal{R}G.$
\item[(e)] $|I||\mathtt{Ann}_{r}(I)|=\mathcal{R}G,$ for a left ideal $I$ in $\mathcal{R}G.$
\item[(f)] $|J||\mathtt{Ann}_{l}(J)|=\mathcal{R}G,$ for a right ideal $J$ in $\mathcal{R}G.$
\end{enumerate}
\end{prop}

\begin{defn}[Checkable codes]$ $ 
\begin{enumerate}\label{t2}
\item[(a)] A right ideal $I$ in $\mathcal{R}G$ is called checkable if there exists an element $u$ in $\mathcal{R}G$ such that $I=\{a~|~a\in\mathcal{R}G,~ua=0\}=\mathtt{Ann}_{r}(u)=\mathtt{Ann}_{r}(\mathcal{R}Gu).$
\item[(b)] A group algebra $\mathcal{R}G$ is called code-checkable if all right ideals of $\mathcal{R}G$ are checkable.
\end{enumerate}
\end{defn}
\begin{thm}\label{t3}
In $\mathcal{R}G$ a right (\textit{resp.} left) ideal $I$ is checkable if and only if $\mathtt{Ann}_{l}(I)$  $(\textit{resp.}~\mathtt{Ann}_{r}(I))$ is a principal left (\textit{resp.} right) ideal. 
\end{thm}
\begin{lemma}\label{lm1}
If $C$ is a right (\textit{resp.} left) ideal in $\mathcal{R}G,$ then $C^{\perp}=\widehat{\mathtt{Ann}_{l}(C)}$ $(\textit{resp.}~ C^{\perp}=\widehat{\mathtt{Ann}_{r}(C)}).$
\end{lemma}
\begin{proof} Let us assume $C$ is a right ideal in $\mathcal{R}G.$
For any element $a\in \widehat{\mathtt{Ann}_{l}(C)},$ this implies $\widehat{a}\in \mathtt{Ann}_{l}(C),$ and $\widehat{a}c g=0$ for all $c\in C$ and $g\in G.$ Then we obtain that $\langle a, c\rangle=0$ for all $c\in C,$ hence $a\in C^\perp.$ Therefore, $\widehat{\mathtt{Ann}_{r}(C)})\subseteq C^{\perp}.$
\begin{align*}
 \mid C^{\perp}\mid& =\dfrac{\mid \mathcal{R} G\mid}{\mid  C\mid} =\mid\widehat{\mathtt{Ann}_{l}(C)}\mid, 
\end{align*}
which completes the proof.
\end{proof}
\begin{thm}\label{Th1.1}
If $C$ is a right ideal (or right $\mathcal{R}G$-submodule) in $\mathcal{R}G$. Then the following statements are equivalent
\begin{itemize}
\item[(1)] $C$ is checkable;
\item[(2)] $C^\perp$ is a principal right ideal.
\end{itemize}
\end{thm}
\begin{proof}
Let us assume $C$ is checkable, thus by Theorem \ref{t2}, $\mathtt{Ann}_{l}(C)$ is a principal left ideal in $\mathcal{R}G,$ i.e., $\mathtt{Ann}_{l}(C)=\mathcal{R}Gu,$ for some $u\in \mathcal{R}G.$ Then, applying Lemma \ref{lm1}, we have that $C^\perp=\widehat{\mathtt{Ann}_{l}(C)}=\widehat{\mathcal{R}Gu}=\widehat{u}\mathcal{R}G,$ hence we get the desired result.
\end{proof}
Note that for checkable code $C$, we have $C^\star\cong \mathcal{R}G/v\mathcal{R}G,$ for some $u\in\mathcal{R}G.$ Since, by applying consequence of Theorem \ref{Th1.1}, i.e., $C^\star\cong \mathcal{R}G/C^{\perp}.$

For a finite group $G$ and a finite Frobenius ring $\mathcal{R},$ we recall that the group ring $\mathcal{R}G$ can be written as $\mathcal{R}G=\bigoplus\limits_{i=1}^{s}e_i\mathcal{R}G,$ where $e_i\mathcal{R}G$ is a indecomposable submodule in $\mathcal{R}G,$ $\sum\limits_{i=1}^{s}e_i=1$, $e_i^2=e_i$, for all $i=1,\dots, s$ and $e_ie_j=0$ for all $i\neq j$. Further, if $e_i$ are primitive idempotents in the centre of $\mathcal{R}G,$ then $e_i\mathcal{R}G$ are both side indecomposable submodule of $\mathcal{R}G.$ \\
If $C$ is a group code in $\mathcal{R}G,$ then \begin{equation}\label{eq:decomp}
C=\sum\limits_{i_j=0}^{t}e_{i_j}\mathcal{R}G,\end{equation}
 where $0\leq t \leq s.$ Therefore, $C=\sum\limits_{i_j=0}^{t}e_{i_j}\mathcal{R}G,$ if and only if $C=\mathtt{Ann}_{r}(\mathcal{R}G(1-\sum\limits_{i_j=0}^{t}e_{i_j}))=\mathtt{Ann}_{r}(\mathcal{R}G\sum\limits_{i_k=0}^{r}e_{i_k})$, where $\sum\limits_{i_j=0}^{t}e_{i_j}+\sum\limits_{i_k=0}^{r}e_{i_k}=1,$  and from   Proposition~\ref{t1}(a) it follows that $C=\mathtt{Ann}_{r}(\sum_{i_k=0}^{r}e_{i_k})$. 

\begin{prop}
	Let $C$ is a group code in $\mathcal{R}G,$  and   $C=\sum\limits_{i_j=0}^{t}e_{i_j}\mathcal{R}G$ its decomposition as in Equation~\ref{eq:decomp} where $\sum\limits_{i_j=0}^{t}e_{i_j}+\sum\limits_{i_k=0}^{r}e_{i_k}=1$. Then $C$ is a checkable code if and only if $C=\bigcap\limits_{i_k=0}^{r}\mathtt{Ann}_{r}(\mathcal{R}Ge_{i_k}).$
\end{prop}
\begin{proof} It follows from the above reasonig and the   well known fact that $$\mathtt{Ann}_{r}(\sum\limits_{i_k=0}^{r}e_{i_k})=\bigcap\limits_{i_k=0}^{r}\mathtt{Ann}_{r}(e_{i_k})=\bigcap\limits_{i_k=0}^{r}\mathtt{Ann}_{r}(\mathcal{R}Ge_{i_k}).$$
	\end{proof}

%%%%%%%%%%%%%%%%%%%%%%%%%%%%%%%%%%%%%%%%%%%%%%%%%%%%%%%%%%%%%%%%%%%%%%%%%%%%%%%%%%%%%%%%%%%%%%%%%%%%%%%%%%%
\section*{Acknowledgements}
 {Third author was partially supported by Grant TED2021-130358B-I00 funded by MCIN/AEI/10.13039/501100011033 and by the “European Union NextGenerationEU/PRTR”}

%%%%%%%%%%%%%%%%%%%%%%%%%%%%%%%%%%%%%%%%%%%%%%%%%%%%%%%%%%%%%%%%%%%%%%%%%%%%%%%%%%%%%%%%%%%%%%%%%%%%%%%%%%%

 \bibliographystyle{plain}
\bibliography{main.bib}

%\begin{thebibliography}{99}

\end{document}